# Automatic scanning of nuclear emulsions with wide-angle acceptance for nuclear fragment detection

**T. Fukuda,**[a,*] **S. Fukunaga,**[a] **H. Ishida,**[a] **K. Kodama,**[b] **T. Matsuo,**[a] **S. Mikado,**[c] **S. Ogawa,**[a] **H. Shibuya,**[a] **and J. Sudo**[a]

[a] *Toho University, Miyama, Funabashi 274-8510, Japan*
[b] *Aichi University of Education, Kariya 448-8542, Japan*
[c] *Nihon University, Narashino 275-8576, Japan*

*E-mail*: `tsutomu.fukuda@ph.sci.toho-u.ac.jp`

ABSTRACT: Nuclear emulsion, a tracking detector with sub-micron position resolution, has played a successful role in the field of particle physics and the analysis speed has been substantially improved by the development of automated scanning systems. This paper describes a newly developed automated scanning system and its application to the analysis of nuclear fragments emitted almost isotropically in nuclear evaporation. This system is able to recognize tracks of nuclear fragments up to $|\tan\theta| < 3.0$ (where $\theta$ is the track angle with respect to the perpendicular to the emulsion film), while existing systems have an angular acceptance limited to $|\tan\theta| < 0.6$. The automatic scanning for such a large angle track in nuclear emulsion is the first trial. Furthermore the track recognition algorithm is performed by a powerful Graphics Processing Unit (GPU) for the first time. This GPU has a sufficient computing power to process large area scanning data with a wide angular acceptance and enough flexibility to allow the tuning of the recognition algorithm. This new system will in particular be applied in the framework of the OPERA experiment : the background in the sample of τ decay candidates due to hadronic interactions will be reduced by a better detection of the emitted nuclear fragments.

---

[*] Corresponding author.

# Contents

## 1. Introduction

Nuclear emulsion is a three-dimensional solid-state tracking detector made of AgBr crystals interspersed in a gelatin matrix. A charged particle passing through an emulsion layer ionizes the crystals along its path and produces latent images. The particle trajectory is visible as a line of metallic silver grains after development. This trajectory is measured with sub-micron position accuracy by using an optical microscope. Therefore the nuclear emulsion is well suited to observe short-lived particles and to measure precisely positions and angles of tracks in the vicinity of the interaction vertex. In fact, the nuclear emulsion technology contributed to the discovery of the pion by C. Powell et al [1] in 1947 and of the charmed particle by K. Niu et al [2] in 1971 and to the understanding of many important phenomena [3]-[6] by human eye-based analysis in the last century.

     Nuclear fragments are produced in the nuclear evaporation process caused by an excitation of the target nucleus when it interacts with an incident particle, such as a neutrino, a hadron or a nucleus (Fig.1 (a)). As the nuclear fragments are emitted almost isotropically, a tracking detector with a large angular acceptance is required to detect them. Nuclear emulsion is sensitive in all directions and thus is suitable for on the study of this process. Therefore the comparative study between the experimental data investigated by human eye in nuclear emulsions and the simulation codes has been made for understanding this process in the past [7][8].

An automated emulsion scanning system, called Track Selector (TS), was designed in the 1970's and has improved the speed of the emulsion analysis dramatically [9]-[11]. This technology was used for the direct observation of tau neutrino interactions by K. Niwa et al [12] in 2000 and for the collection of a large sample of charmed particle decays in the CHORUS experiment [13]. The latest versions [14][15] of this scanning system were developed for the OPERA experiment [16][17] aiming to detect $\nu_\tau$ appearance in the CNGS $\nu_\mu$ beam. In these applications, most of the tracks of interest are rather collimated with the beam direction and the scanning system has thus been optimized to be fast and efficient for tracks with $|\tan\theta| \leq 0.6$ (Fig.1 (b)). To detect nuclear fragments from interactions in the OPERA experiment, we developed a new automatic scanning system having a much wider angular acceptance of $|\tan\theta| \leq 3.0$. This paper describes its performance and presents some of its possible applications.

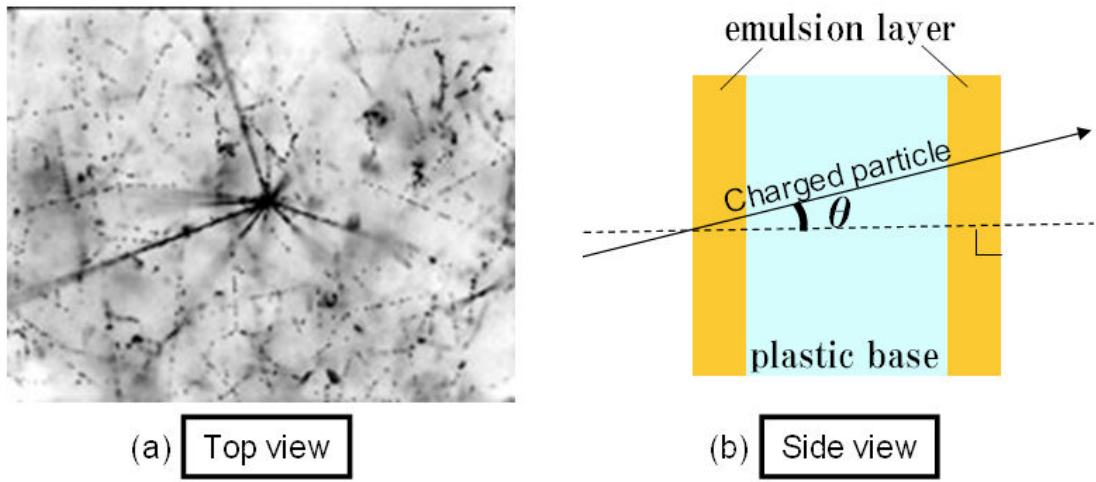

Figure 1. (a) Microscope view of an emulsion film. Some nuclear fragments are emitted from an interaction vertex. (b) Cross-sectional view of an emulsion film. The angle θ of a charged particle track is taken with respect to the perpendicular to the film (i.e. almost the incident beam direction in this study).

## 2. New scanning system

### 2.1 Track recognition

The algorithm of track recognition in TS was devised by K. Niwa [9] and the practical use of TS in particle physics was accomplished by S. Aoki [10] and T. Nakano [11]. A track recognition algorithm of our new system is in principle the same as in TS but tuned for large angle tracks. The nuclear emulsion films used in this study are OPERA films [18] in which 44 μm emulsion layers are coated on both sides of a 205 μm thick plastic base. The OPERA films were mass-produced by Fujifilm Corporation and are being used also in many other experiments. The diameter of the metallic silver grains after development is about 0.6 μm. The optics must be chosen such that the effective pixel size remains small compared to the grain size for a sensor with a given number of pixels. Therefore one pixel size is generally adjusted to be about 0.3 μm.

The track search is performed independently for each emulsion layer and for each field of view. The different steps of the track recognition algorithm are described below.

(1) Using a CMOS image sensor through the microscope optics, 16 tomographic images which have brightness information on each pixel are taken from an emulsion layer (Fig.2 (a)).

(2) All the tomographic images are processed by a convolution filter for smoothing (Fig.2 (b)).

(3) Brightness information of each pixel in these images is binarized applying a threshold level (Fig.2 (c)).

(4) Hit pixels are expanded in both X and Y directions to control grain sizes which affect track recognition efficiency (Fig.2 (d), Fig.3).

(5) 16 processed images or a list of hit pixels are investigated to take coincidence in every angle within set acceptance to recognize a series of grains on a straight line as a track (Fig.2 (e)).

(6) If the sum of the number of hit layers of a track, called Pulse Height (PH), is above a set threshold, it is read out as a signal (Fig.2 (f)).

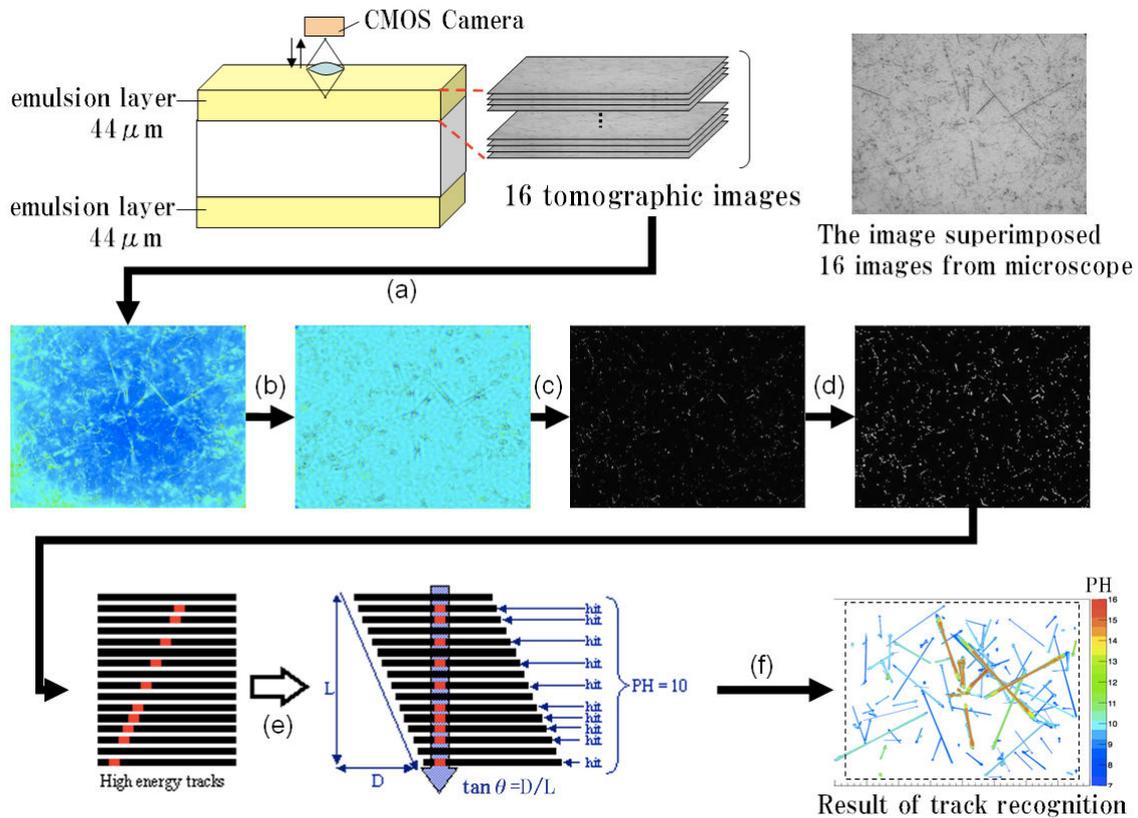

Figure 2. The track recognition algorithm; (a) image data taking, (b)-(d) image processing, (e)-(f) straight track reconstruction. At first, there are various information data such as (x, y, z, brightness) for all pixels. These information data are compressed through the all track recognition processes. Finally, they provide position (x, y), angle (tan $\theta$ x, tan $\theta$ y) and PH information for each recognized track.

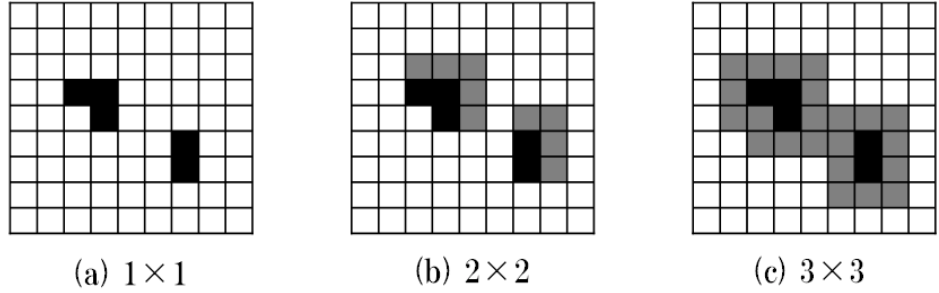

Figure 3. The expansion process. The detail of Fig.2 (d) is described. Original hit pixels (black) are shown in (a). Gray pixels are added by the expansion process in (b) and (c). Both black and gray pixels are used to recognize the tracks. The grain size for track recognition is controlled by this process. Larger grain size improves the efficiency of signal tracks, while the angular resolution becomes worse and more fake tracks emerge. So the choice of an expansion filter depends on various conditions. A 3×3 expansion filter is usually used for the track recognition of minimum ionizing particles with conventional angular acceptance [14].

The superposition of the images need to be performed for each hypothesis on the track angle ($\theta$ and $\phi$) within acceptance. Thus, the processing time will increase strongly with the angular acceptance.

**2.2 Setup for recognizing large angle tracks**

The track recognition process in the scanning system is performed by each field of view. The step size of microscope movements must be chosen such that the tracks of interest have a high probability to be contained in at least one field of view. But if the step size is too small, the scanning time become too long. One emulsion layer has a thickness of 44 μm in the case of the OPERA film, thus a track with $|\tan\theta|=3.0$ extends over 132 μm in the microscope view. Assuming the side of the microscope view to be $L_{VIEW}$, one can move a view by $L_{STEP}=L_{VIEW}$ - 132 μm to scan an emulsion layer without losing any tracks with $|\tan\theta|<3.0$. This means that $L_{VIEW}$ must be substantially larger than 132 μm and a larger microscope view is needed for efficient scanning with wider angle acceptance. Therefore camera and objective lens are newly selected for the system.

Mikrotron Eosens MC 1362 is used as a CMOS image sensor. The resolution of this sensor is 1280 pixels × 1024 pixels and its frame rate is 506 fps. Nikon CFI plan oil immersion lens is used as an objective lens. It has a 50 times magnification, a numerical aperture of 0.90 and a working distance of 0.35 mm. The field of view is set to be 352 μm × 282 μm corresponding to a pixel size of 0.28 μm. The large computing power needed to process the track recognition algorithm over a wide angular acceptance is provided by the Graphics Processing Unit (GPU). The GPU used here is NVIDIA Tesla C2050, which has 448 cores with 3 GB memory and a peak single precision performance of 1.03 Tflops.

Photographs of the new system are shown in Fig.4. The microscope stage, on which the new camera and optics are mounted, is a conventional one.

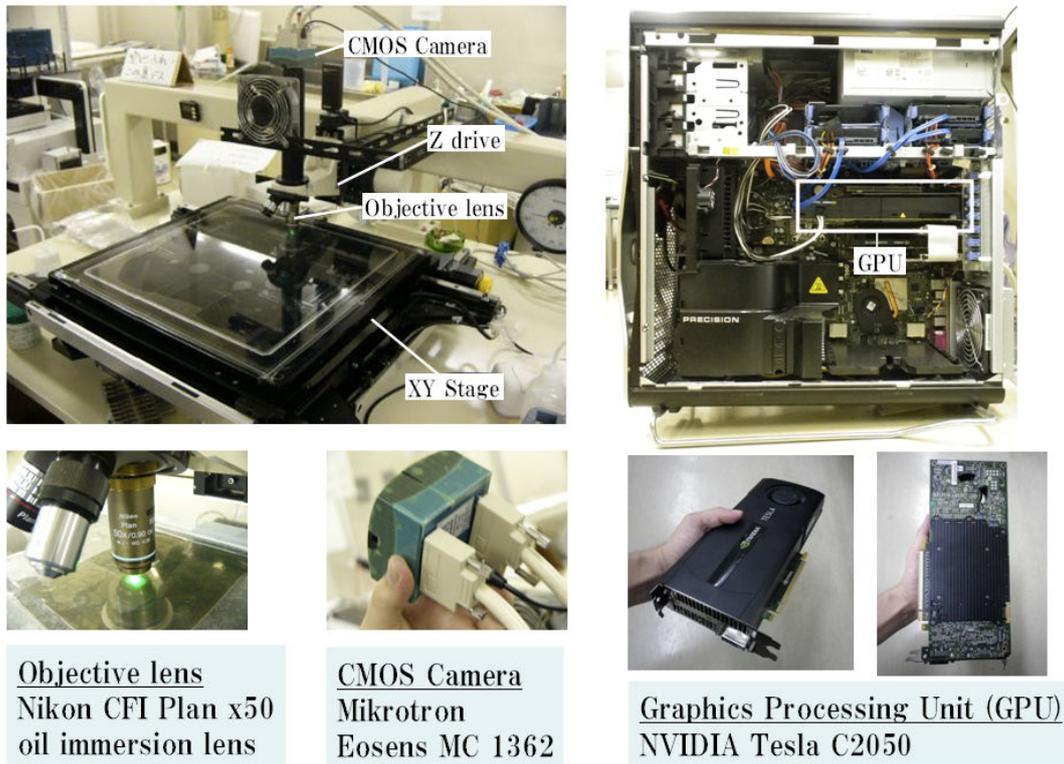

Figure 4. Photographs of the new scanning system with wide-angle acceptance.

## 3. Performance evaluation

The performance of the new scanning system has been evaluated using a stack of emulsion films exposed to a high energy pion beam. Details on the emulsion films and their exposure are given in Section 3.1 and an example of track data obtained by the automatic scanning is described in Section 3.2. The method to tune the scanning parameters is described in Section 3.3. The result on track recognition efficiency and angular measurement accuracy are shown in Section 3.4 while the processing speed is shown in Section 3.5.

### 3.1 Exposure of the emulsion films

An Emulsion Cloud Chamber (ECC) consisting of 30 OPERA films interleaved with 1mm-thick lead plates, was exposed to a pion beam described in Table 1. Most of the pion interactions occurred in the lead plates, as illustrated in Fig.5.

Table 1. Beam conditions

| Beam line | CERN-PS T7 |
|---|---|
| Particle | $\pi^-$ |
| Momentum | 10 GeV/c |
| Density | $5 \times 10^3$ particles/cm$^2$ |

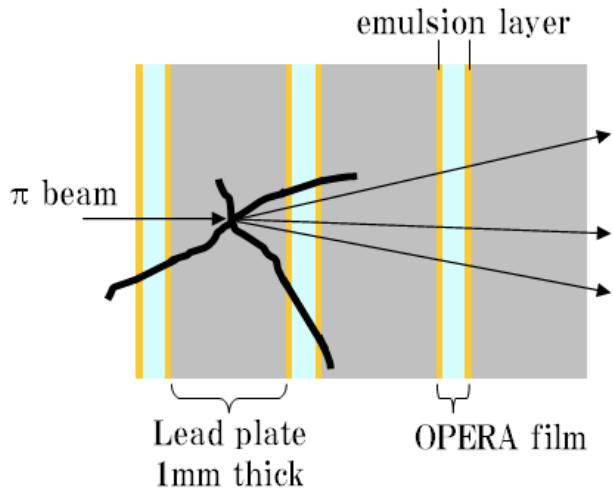

Figure 5. Schematic view of a π interaction in a lead plate. High energy particles (thin arrow) are emitted to forward and nuclear fragments (heavy line) are emitted almost isotropically from the interaction vertex. The tracks of these charged particles are only recorded in the sensitive emulsion layers.

## 3.2 Automatic scanning

Fig.6 shows the result of a 5.0 mm × 3.5 mm area scanning with wide-angle acceptance in a film. All recognized tracks are shown as arrows with a length representing for their slope and for a color representing their PH value. The whole area is shown in the left and a 1 mm side region is drawn enlarged in the right. A lot of tracks of sub-MeV or a few MeV electrons as shown in Fig.7 (a) from environment are recorded in emulsion, because nuclear emulsion starts to record all tracks immediately after it is produced and before it is developed with no dead time. Therefore the automatic scanning system recognizes many "fake" tracks made of low energy electrons and/or random noise grains, called fogs. Most of recognized tracks are of such kinds. They consist of a straight part in curved low energy electron track due to multiple Coulomb scattering and/or chance coincidences of fogs in emulsion. Since the PH value of such "fake" tracks are small, PH distribution has concentrated at small PH value as shown typically in Fig.7 (b).

A particle, penetrated an emulsion film, creates tracks on both emulsion layers. These two tracks are connected to form a corresponding base track with a slope defined by the line connecting intersection points of the tracks in both emulsion layers with the surface of the 205μm thick plastic base. Then base tracks are also connected across lead plates to reconstruct tracks over the whole chamber. In this process, most of "fake" tracks are eliminated [19][20].

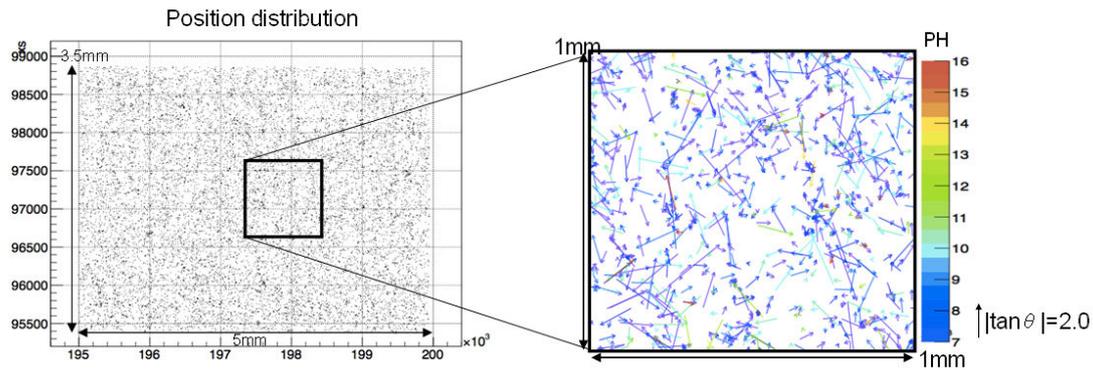

Figure 6. Distribution of the readout tracks in the 5.0 mm × 3.5 mm scanned area.

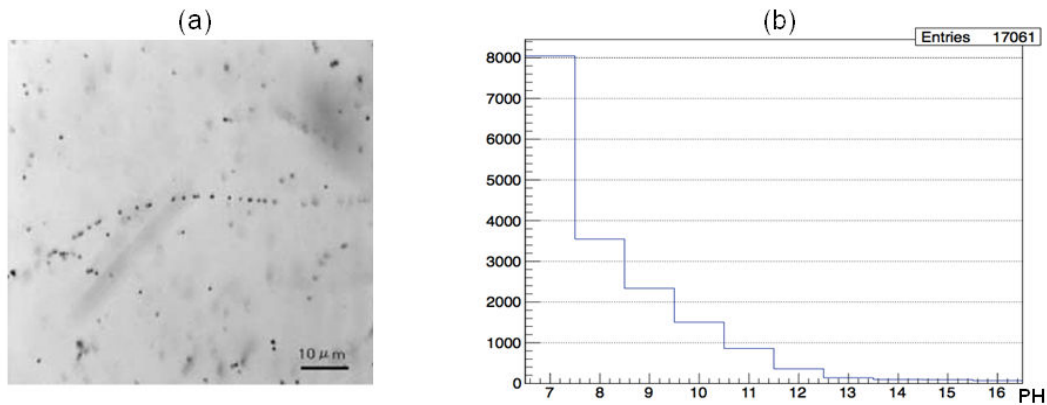

Figure 7. (a) a microscope image of a curved track of an electron of about 1 MeV. (b) PH distribution of readout tracks in the 5.0 mm × 3.5 mm scanned area.

### 3.3 Tuning of the scanning parameters using small angle MIPs

The scanning parameters (the threshold of brightness, the expansion filter, the threshold of PH) are tuned by using minimum ionizing particles (MIPs) in conventional angle acceptance |tan $\theta$ | < 0.6. When a MIP passes through an emulsion layer of an OPERA film, about 30-35 grains per 100μm are created by ionization after development. Hence the number of grains in a 44μm thick emulsion layer is distributed as a binominal distribution, with a mean value of about 15. The peak value of the PH distribution is typically 10-11 in our current scanning system mainly due to the focal depth of the optics.

A set of MIP tracks reconstructed by our current system is used as a reference to tune the scanning parameters of the new system. The scanning parameters in the new system were tuned to reproduce the performance of the current system : a 3×3 expansion filter is used and the PH threshold level is fixed at 7 and brightness threshold is tuned properly. The PH distributions for MIPs after the tuning are shown in Fig.8 for different angular intervals.

The angular accuracy of MIP tracks scanned by the new system was also measured. It is evaluated from the angle difference between the track recognized in a layer and the corresponding base track. The angular accuracy is 14.5 mrad for $0.0 \leq |\tan \theta| < 0.2$, 16.9 mrad for $0.2 \leq |\tan \theta| < 0.4$, and 24.4 mrad for $0.4 \leq |\tan \theta| < 0.6$ as shown in Fig.9. They are almost the same as for our current scanning system [14]. The slight worsening of the accuracy with increasing track angle is a general feature of automatic scanning systems.

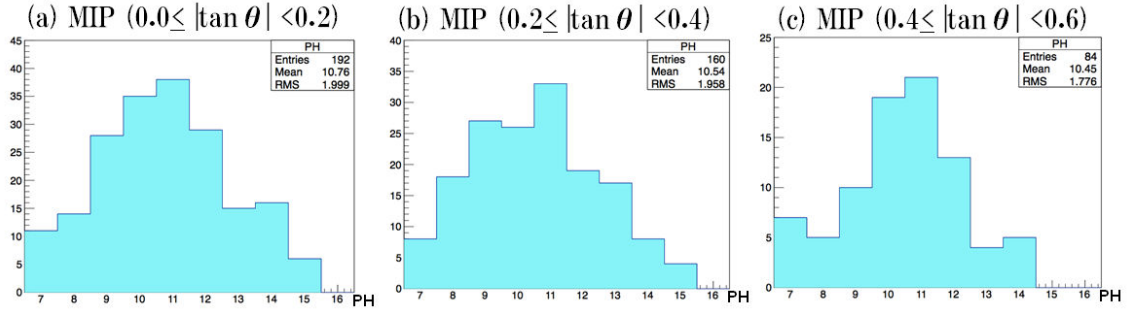

Figure 8. PH distribution of MIP tracks in different angular intervals within the conventional acceptance. (a) is for $0.0 \leq |\tan \theta| < 0.2$; (b) is for $0.2 \leq |\tan \theta| < 0.4$; (c) is for $0.4 \leq |\tan \theta| < 0.6$.

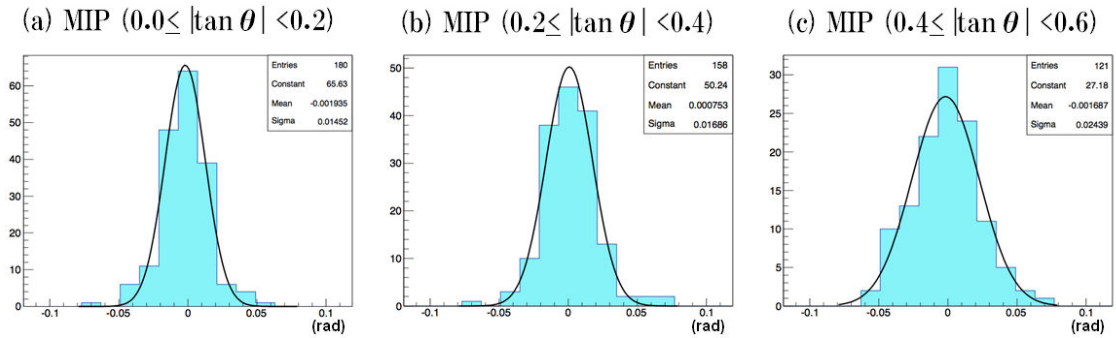

Figure 9. Angular accuracy of MIP tracks in different angle intervals within the conventional acceptance. (a) is for $0.0 \leq |\tan \theta| < 0.2$; (b) is for $0.2 \leq |\tan \theta| < 0.4$; (c) is for $0.4 \leq |\tan \theta| < 0.6$.

**3.4 Scanning result for nuclear fragments in wide-angle acceptance**

Tracks of highly ionizing nuclear fragments are recorded as thick heavy lines in emulsion and are called black tracks. They are easily identified by human eyes as shown in Fig.1 (a). A sample of nuclear fragment tracks, which penetrate in emulsion layer, was first picked up by eyes and then scanned by the new scanning system with PH threshold = 7. The number of picked up nuclear fragment tracks is 91 for $0.0 \leq |\tan \theta| < 1.0$, 76 for $1.0 \leq |\tan \theta| < 2.0$ and 70 for $2.0 \leq |\tan \theta| < 3.0$. All these tracks were successfully recognized by the new system and their PH distribution in each angle region is shown in Fig.10. The mean tracking efficiency for nuclear fragments ($2.0 \leq |\tan \theta| < 3.0$) is thus larger than 99.8 % at 90 % confidence level when the PH threshold is 13. Therefore the PH threshold for nuclear fragments could be set at 12 or 13 without loss of tracking efficiency.

Angular accuracy of nuclear fragment tracks which penetrate an OPERA film is evaluated using the same method as of MIPs described in Section 3.3. As shown in Fig.11, angular accuracy is 14 mrad for $0.0 \leq |\tan \theta | < 1.0$, 21 mrad for $1.0 \leq |\tan \theta | < 2.0$ and 33 mrad for $2.0 \leq |\tan \theta | < 3.0$. Angular accuracy of black track also depends on its angle like in the case of small angle MIPs.

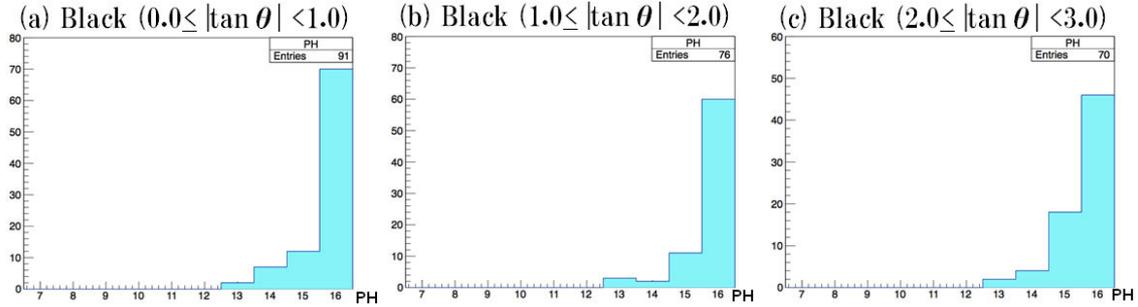

Figure 10. PH distribution of nuclear fragment tracks in each angle. (a) is for $0.0 \leq |\tan \theta |< 1.0$; (b) is for $1.0 \leq |\tan \theta |< 2.0$; (c) is for $2.0 \leq |\tan \theta |< 3.0$.

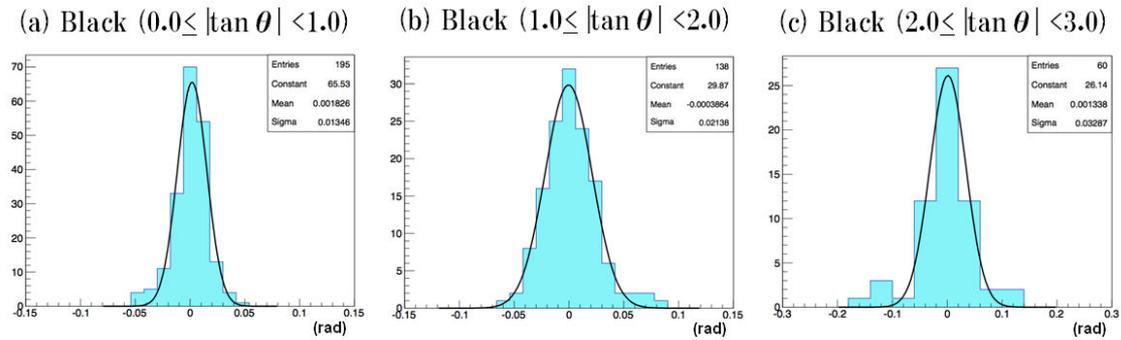

Figure 11. The angular accuracy of nuclear fragment tracks in different angular intervals. (a) is for $0.0 \leq |\tan \theta |< 1.0$; (b) is for $1.0 \leq |\tan \theta |< 2.0$; (c) is for $2.0 \leq |\tan \theta |< 3.0$.

### 3.5 Processing speed for track recognition

The whole track recognition algorithm described in Section 2.1 is processed using the GPU of the new scanning system. The GPU was intended for graphics display in the past, but is currently also used for general purpose high performance computing, taking advantage of its large number of computing cores. The Integrated Development Environment for GPU programming used here is CUDA 4.0.

Fig.12 shows the measured processing time per view for conventional angle acceptance ($|\tan \theta |\leq 0.6$) and for wide angle acceptance ($|\tan \theta |\leq 3.0$), using parameters tuned as described in previous sections. They are about 100 msec and about 550 msec respectively. This processing speed is acceptable for actual application. The processing time could be 2-3 times faster by replacing the present GPU by the latest version hardware (NVIDIA GTX680).

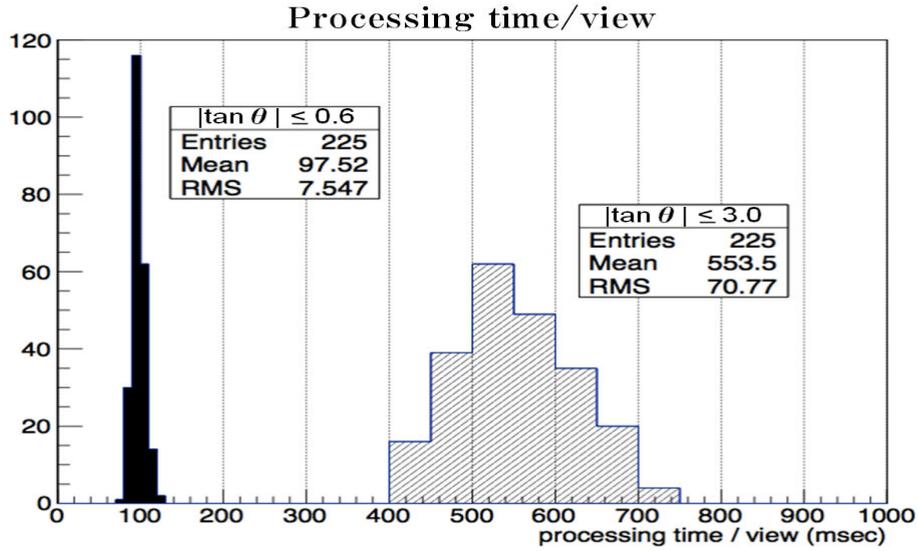

Figure 12. Black and dashed histograms of the processing time in conventional and wide-angle acceptance.

## 4. Discussion for applications

Basic performance of the newly developed automatic emulsion scanning system to recognize nuclear fragments with wide angle acceptance is described here. Track recognition efficiency and angular accuracy of nuclear fragment tracks are enough for physics analysis.

In the OPERA experiment, it is one of the main background to $\tau$ decay in $\nu_\tau$CC interaction when hadrons from $\nu_\mu$ NC interactions interact in a lead plate immediately after. These backgrounds can be reduced by detecting nuclear fragments from a second vertex, since the existence of nuclear fragments is clear proof of a hadron interaction, not a $\tau$ decay. This new scanning system is planned to be applied to detect nuclear fragments from second vertices to identify and to eliminate hadron interactions from $\nu_\tau$CC interaction candidate events in the OPERA experiment.

This scanning system is also a good tool for detail and systematic analysis of nuclear evaporation process from neutrino, hadron and stable/unstable - nucleus interactions. The emulsion gel itself, i.e. the mixture material was used as the target material for the analysis of this process in the past experiments [7][8]. Hence the target nucleus was Ag, Br, C, O, N, H, I, S and so on. If the ECC type target was used, it allows us to investigate this process in a variety of pure materials by changing the target materials inserted between emulsion films. Though large area scanning is required to search large angle nuclear fragments from many interactions because the interaction points are in the target material and large angle tracks run over a long distance to reach emulsion films, this system will make it possible to search large angle tracks effectively. In the case of such analysis, it should be considered that the energy threshold of detectable nuclear fragments depends on their slope and the depth of the interaction vertex in the target material.

It will also be possible to measure MIPs with wide angle acceptance by tuning parameters of this system properly, which will open possibilities for nuclear emulsion to be applied to many other applications.

## 5. Conclusions

In this paper, first results of track recognition for nuclear fragments with wide angle acceptance, using the new automatic emulsion scanning system, is described. Track recognition in the new system is confirmed to show the good quality although angle acceptance is 5 times larger than our current system.

The GPU is successfully applied in all processes of track recognition, i.e. from image smoothing to track selection, for the first time. This allows systematic analysis in nuclear emulsions with wide angle acceptance efficiently.

The new system will first be applied to background reduction in the OPERA experiment and also to detail analysis of nuclear evaporation process from interactions of neutrino, hadron and stable/unstable nucleus.

## 6. Acknowledgments

We appreciate the support provided by the Toho University and the collaborating laboratories. We also express special thanks to M. Kimura and A. M. Ito for the fruitful discussions. We acknowledge the support from the Japan Society for the Promotion of Science (JSPS) through their grants. We also gratefully acknowledge P. Vilain for his careful reading of the manuscript.